\documentclass[twocolumn]{aastex631}

\usepackage{array}
\usepackage{amsmath}
\usepackage{makecell}
\usepackage{multirow}

\newcommand{\IMRELEASENO}{LLNL-JRNL-858470}

\newcommand{\edits}[1]{{\color{black} #1}}

\newcommand{\llnl}{Space Science Institute, Lawrence Livermore National Laboratory, 7000 East Ave., Livermore, CA 94550, USA}
\newcommand{\UCB}{University of California, Berkeley, Astronomy Department, Berkeley, CA 94720, USA}
\newcommand{\UCR}{Department of Physics and Astronomy, University of California, Riverside, 900 University Avenue, Riverside, CA 92521, USA}

\begin{document}

\title{Astrometric Microlensing by Primordial Black Holes with The Roman Space Telescope}

\author[0009-0001-5288-0009]{James Fardeen\,}
\affiliation{\llnl}
\author[0000-0002-1052-6749]{Peter McGill\,*}
\email{*mcgill5@llnl.gov}
\affiliation{\llnl}
\author[0000-0002-5910-3114]{Scott E. Perkins\,}
\affiliation{\llnl}
\author[0000-0003-0248-6123]{William A. Dawson\,}
\affiliation{\llnl}
\author[0000-0002-0287-3783]{Natasha S. Abrams}
\affiliation{\UCB}
\author[0000-0001-9611-0009]{Jessica R.~Lu\,}
\affiliation{\UCB}
\author[0000-0002-4457-890X]{Ming-Feng Ho\,}
\affiliation{\UCR}
\author[0000-0001-5803-5490]{Simeon Bird}
\affiliation{\UCR}

\begin{abstract}

\noindent Primordial Black Holes (PBHs) could explain some fraction of dark matter and shed light on many areas of early-universe physics. 
Despite over half a century of research interest, a PBH population has so far eluded detection. The most competitive constraints on the fraction of dark matter comprised of PBHs ($f_{\rm DM}$) in the $(10^{-9}-10)M_{\odot}$ mass-ranges come from photometric microlensing and bound $f_{\rm DM}\lesssim10^{-2}-10^{-1}$. 
With the advent of the Roman Space Telescope with its sub-milliarcsecond (mas) astrometric capabilities and its planned Galactic Bulge Time Domain Survey (GBTDS), detecting astrometric microlensing signatures will become routine. 
Compared with photometric microlensing, astrometric microlensing signals are sensitive to different lens masses-distance configurations and contains different information, making it a complimentary lensing probe. 
At sub-mas astrometric precision, astrometric microlensing signals are typically detectable at larger lens-source separations than photometric signals, suggesting a microlensing detection channel of pure astrometric events. 
We use a Galactic simulation to predict the number of detectable microlensing events during the GBTDS via this pure astrometric microlensing channel. \edits{Assuming an absolute astrometric precision floor for bright stars of 0.1 mas for the GBTDS, we} find that the number of detectable events peaks at $\approx 10^{3} f_{\rm DM}$ for a population of $ 1 M_{\odot}$ PBHs and tapers to $\approx 10f_{\rm DM}$ and $\approx 100f_{\rm DM}$ at $10^{-4}M_{\odot}$ and $10^{3}M_{\odot}$, respectively. 
Accounting for the distinguishability of PBHs from Stellar lenses, we conclude the GBTDS will be sensitive \edits{to a} PBH population at $f_{\rm DM}$ down to $\approx10^{-1}-10^{-3}$ for $(10^{-1}-10^{2})M_{\odot}$ likely yielding novel PBH constraints.

\end{abstract}

\keywords{}

\section{Introduction} \label{sec:intro}

Primordial Black Holes (PBHs) are theorized to have formed through density fluctuations on the cosmological horizon in the early, radiation-dominated universe \citep{Zeldovich1967,Hawking1971}. 
Among the many implications of the existence of PBHs (e.g., from seeding supermassive black holes in galaxies; \citealt{Kawasaki2012,Bernal2018}, to explaining the Galactic $\gamma-$ray background; \citealt{Carr2016}), they are possible dark matter (DM) candidates \citep{Chapline1975}. Constraints on a PBH population would also provide insights into early-universe physics \citep[e.g.,][]{Carr1975, Bird2023}. 

Despite a diverse set of expected observable signatures and over half a century of research interest, there is still no compelling evidence for the existence of PBHs \citep{Carr2016dm,Carr2020, Green2021}.
Spanning $40$ orders of magnitude in PBH mass, probes ranging from the cosmic microwave background ~\citep[e.g.,][]{Ricotti2007}, to gravitational waves of merging black holes ~\citep[BHs; e.g.,][]{Franciolini2021}, to microlensing \citep[e.g.,][]{Wyrzykowski2011b}, have placed bounds on the fraction of DM explainable by PBHs ($f_{\rm DM}$), but there have been no definitive PBH detections. 

Looking towards the Magellanic Clouds, the original microlensing surveys placed upper bounds on $f_{\rm DM}\approx 1-10\%$ in the mass ranges $(10^{-7}-10^{3}) M_{\odot}$ \citep{Alcock2001,Tisserand2006,Blaineau2022,Wyrzykowski2009,Wyrzykowski2010,Wyrzykowski2011a,Wyrzykowski2011b}. A subsequent high-cadence microlensing survey towards M31 placed upper bounds on $f_{\rm DM}\approx0.1-1\%$ at sub-stellar $(10^{-14}-10^{-9}) M_{\odot}$ mass ranges~\citep{Niikura2019a}. 
Most recently, observation towards the Galactic bulge from the Optical Gravitational Lensing Experiment \citep[OGLE;][]{Udalski2015,Mroz2017} has placed constraints on $f_{\rm DM}\approx1-10\%$ for a mass range of $(10^{-6}-10^{-3})M_{\odot}$ ~\citep{Niikura2019b}, and methods to robustly disentangle PBH and astrophysical lens populations using photometric microlensing surveys towards the Bulge have been developed \citep[e.g.,][]{Perkins2023}.

The aforementioned constraints have solely relied on photometric microlensing signals. 
However, with the advent of space observatories capable of sub-milliarcsecond (mas) astrometry such as Gaia \citep{Prusti2016}, the Hubble Space Telescope (HST), and the Nancy Grace Roman Space Telescope \citep[RST;][]{Spergel2015}, as well as ground-based adaptive optics systems \citep[e.g.,][]{Lu2016, Zurlo2018}, it is also possible to detect astrometric microlensing signals \citep{Walker1995, Hog1995, Miyamoto1995}. 
Although these astrometric and photometric signals arise from the same underlying phenomena, their characteristics and information content differ \citep[e.g.,][]{Dominik2000,Belokurov2002}. 
Notably, unlike the photometric signal, the detection of the astrometric signal can lead to lens-mass determinations \citep{Lu2016,Sahu2017,Zurlo2018,Sahu2022,Lam2022,McGill2023, Lam2023}. 
Overall, photometric and astrometric microlensing signals can offer complementary probes and are sensitive to different lens mass-distance configurations \citep{Dominik2000}.

At sub-mas astrometric precision, astrometric microlensing signals are typically detectable at larger lens-source separations compared with photometric events. 
This makes the astrometric optical depth (probability of lensing) higher than the photometric optical depth \citep{Miralda-Escude1996, Dominik2000}. 
Indeed, it is possible and likely for a microlensing event to happen at a wide enough impact parameter to cause an astrometric signal but no detectable photometric amplification \citep[e.g.,][]{Sahu2017, Zurlo2018, Bramich2018, McGill2018, McGill2020, McGill2023}.

In the wide lens-source separation and purely astrometric regime, \cite{VanTilburg2018} derived theoretical sets of spatially correlated astrometric observables. These lensing signals can act over many sources simultaneously \citep[e.g.,][]{Gaudi2005, DiStefano2008}, can be caused by sub-halo dark matter, and are expected to be detectable by Gaia \citep{Mishra-Sharma2020, Chen2023, Mondino2023}. 
Motivated by Gaia's unprecedented all-sky astrometric survey, \cite{Verma2023} predicted that Gaia will be sensitive to PBHs via astrometric microlensing in the range of $(0.4- 5\times10^{7}) M_{\odot}$ probing to $f_{\rm DM}\approx3\times10^{-4}$. 
\cite{VanTilburg2018} and \cite{Verma2023} highlighted the power of astrometric microlensing as a probe of dark matter in the sub-mas astrometry era.

Due to be launched in the late 2020s, RST will have a similar resolution and wavelength coverage to HST but with $100$ times the field of view\footnote{\url{https://roman.gsfc.nasa.gov/}}.
RST will carry out the Galactic Bulge Time-Domain Survey \citep[GBTDS;][]{Penny2019} with one of its main goals to conduct a census of planets in the Galaxy via photometric microlensing \citep[e.g.,][]{Bennett2002, Penny2019,Johnson2020}. 
The GBTDS will survey $\approx2$ deg$^{2}$ over 5 years in the infrared wavelengths. 
In addition to benefiting from increased microlensing event rates in the infrared \citep[e.g.,][]{Gould1994, McGill2019, Husseiniova2021, Kaczmarek2022, Kaczmarek2023, Luberto2022, Wen2023, Kondo2023}, the GBTDS will take simultaneous photometry and astrometry at a $15$ minute cadence during an observing season. 

High-cadence astrometry and photometry over regions of high microlensing optical depth presents an unprecedented opportunity for detecting isolated objects. Recent work has focused on exploiting high-cadence photometry or joint photometry and astrometry during the GBTDS to characterize both isolated black holes \citep{Sajadian2023, Lam2023b,Mroz2022} and PBHs \citep{Pruett2022, DeRocco2023}. 
However, the GBTDS's sensitivity to PBHs in the wide-separation, purely astrometric microlensing regime has yet to be investigated.  

In this work, we use a Galactic simulation to investigate the GBTDS's sensitivity to detecting PBHs lenses purely astrometrically with masses ranging ($10^{-4}-10^{3}$)$M_{\odot}$. 
In Section \ref{sec:astrometric_microlensing}, we review relevant astrometric microlensing characteristics. 
In Section \ref{sec:simulation}, we describe our simulation and methods to extract astrometric microlensing events. 
In Section \ref{sec:results}, we calculate predicted numbers of detectable PBH microlensing events, assess PBH lens distinguishability from the astrophysical lens population of Stars, Neutron Stars (NSs), White Dwarfs (WDs), and stellar-origin Black Holes (SOBHs) and derive the GBTDS's sensitivity in the context of current PBH constraints. 
Finally, in Section \ref{sec:conclusions} we summarize our findings discuss further implication of this work. 

\section{Astrometric Microlensing}
\label{sec:astrometric_microlensing}

\noindent Microlensing occurs during the alignment of a background source (at distance $D_{S}$) and intervening lensing object, of mass $M_{L}$ at distance $D_{L}$, where $D_{L}<D_{S}$ are distances from an observer.
Under perfect lens-source alignment, an Einstein ring image of the source is formed with angular radius \citep{Einstein1936},
\begin{equation}
\theta_\mathrm{E} = \sqrt{\frac{4GM}{c^2}\frac{D_\mathrm{S} - D_\mathrm{L}}{D_\mathrm{S}D_\mathrm{L}}}\,.
\label{eq:theta_E}
\end{equation}
In the case of imperfect lens-source alignment, two source images are formed \citep[e.g.,][]{Paczynski1986largeopticaldepth},
\begin{equation}
    \boldsymbol{\theta_{\pm}}(\boldsymbol{u}) = \left(|\boldsymbol{u}|\pm\sqrt{|\boldsymbol{u}|^{2}+4}\right)\frac{\theta_{E}\hat{\boldsymbol{u}}}{2}\,.
    \label{eq:image_positions}
\end{equation}
$\boldsymbol{u}$ being the lens-source angular separation (positive direction towards the source position), normalized by $\theta_{E}$, and $\hat{\boldsymbol{u}}=\boldsymbol{u}/|\boldsymbol{u}|$. 
As a function of time, $t$,
\begin{equation}
    \boldsymbol{u}(t) = \boldsymbol{u_{0}} + \frac{(t - t_{0})}{t_{E}}\boldsymbol{\hat{\mu}_{\text{rel}}} + \boldsymbol{P}(t; \boldsymbol{\pi_{E}})\,.
\end{equation}
Here, $t_{0}$ is the time of lens-source closest approach, $\boldsymbol{u_{0}}$, and hereafter $|\boldsymbol{u_{0}}|=u_{0}$. $\boldsymbol{\hat{\mu}_{\text{rel}}}$ is the relative lens-source proper motion unit vector, $\boldsymbol{P}$ is the lens-source parallax motion, $t_{E}$ and $\boldsymbol{\pi_{E}}$ are the Einstein timescale and microlensing parallax given by,

\begin{figure*}[t!]
    \centering
    \includegraphics[width=\textwidth]{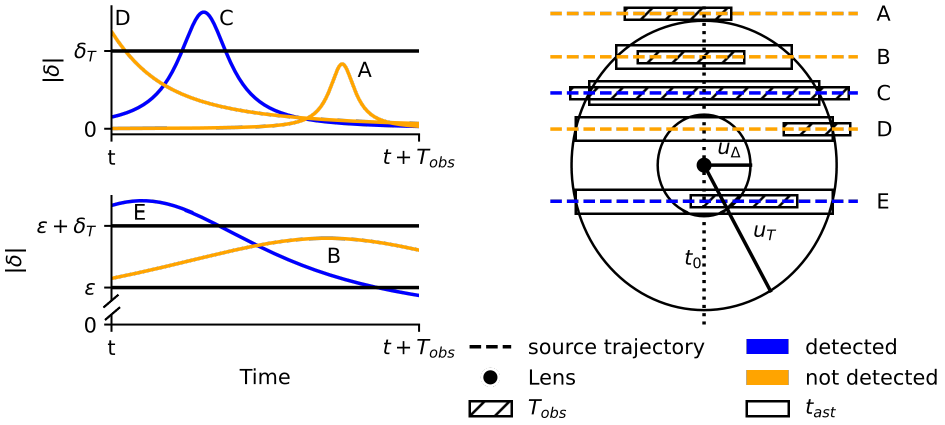}
    \caption{A \edits{schematic} of example astrometric microlensing events ($u_{0}>2$) and detection criteria. \edits{The left plots show the magnitude of the astrometric signal for a selection of events, and the right plot shows their corresponding source trajectories. In the left plots, $\varepsilon>>\delta_{T}$ is an arbitrary astrometric offset chosen to show events with a change in astrometric signal above the threshold ($\delta_{T} + \varepsilon - \varepsilon = \delta_{T}$) within $T_{\rm obs}$. Given some observational window of length $T_{\rm obs}$, the astrometric signal for each geometry is only detectable when it exceeds $\delta_t$ over a timespan of $t_{\rm ast}$. In the right plot, the rectangles show the distance travelled by the source in a time period $T_{\rm obs}$ ({\em hashed box}) and $t_{\rm ast}$ ({\em open box}) in the rest frame of the lens. Their vertical heights differ so they can be easily visually distinguished. Event A is not detected because $u_0>u_{T}$. Event B is not detected because although $u_{0}<u_{T}$, $t_{\rm ast}>T_{\rm obs}$ and the event does not have a detectable change in astrometric signal within $T_{\rm obs}$, i.e. $u_{0}>u_{\Delta}$. Event C is detected because $t_{\rm ast}< T_{\rm obs}$, $u_{0}<u_{T}$ and the event peaks within $T_{\rm obs}$. Event D is not detected because it does not peak within $T_{\rm obs}$, although $u_{0}<u_{T}$. Event E is detected because in spite of $t_{\rm ast}>T_{\rm obs}$, $u_{0}< u_{\Delta}$, and the event peaks within $T_{\rm obs}$ - i.e., Event E has a sufficiently large change in the astrometric signal within $T_{\rm obs}$.}
    }
    \label{fig:ast_signals}
\end{figure*}
\begin{equation}
    t_{E} = \frac{\theta_{E}}{\mu_{\text{rel}}},\quad{}\quad{}|\boldsymbol{\pi_{E}}|=\frac{1}{\theta_{E}}\left(\frac{1}{D_{L}}-\frac{1}{D_{S}}\right),
    \label{eq:tE_piE}
\end{equation}
respectively. 
$\boldsymbol{\pi_{E}}$ is parameterized with components in the north ($\pi_{EN}$) and east ($\pi_{EE}$) directions. 
During the event, the source images change position and brightness causing both photometric \citep{Paczynski1986largeopticaldepth} and astrometric effects \citep{Walker1995, Hog1995, Miyamoto1995}. Assuming a dark lens and no blended light, the astrometric shift due to microlensing from the unlensed source position is \citep[e.g.,][]{Bramich2018}, 
\begin{equation}
\boldsymbol{\delta}(\boldsymbol{u}) = \frac{\boldsymbol{u}}{|\boldsymbol{u}|^{2}+2}\theta_{E}.
\label{eq:full_shift}
\end{equation}
$\boldsymbol{\delta}$ is scaled by $\theta_{E}$ and has a maximum amplitude of $\approx0.354\theta_{E}$ at $|\boldsymbol{u}|=\sqrt{2}$ or at $u_{0}$ for events with $u_{0}>\sqrt{2}$ differing from the characteristics of photometric microlensing. 
In photometric microlensing, the amplification can increase almost arbitrarily for arbitrarily close lens-source impact parameters and is only eventually bounded by finite-source effects \citep[e.g.,][]{Rybicki2022}. 
At $|\boldsymbol{u}|>>\sqrt{2}$, Eq.~\eqref{eq:full_shift} is approximately \citep{Dominik2000},
\begin{equation}
    \boldsymbol{\delta}(\boldsymbol{u}) \approx \frac{\theta_{E}}{|\boldsymbol{u}|}\hat{\boldsymbol{u}}\,.
    \label{eq:shift}
\end{equation}
This approximation overestimates $|\boldsymbol{\delta}|$. Hereafter we denote $|\boldsymbol{\delta}|=\delta$. Assuming some astrometric detection threshold, $\delta_{T}$, Eq.~\eqref{eq:shift} can be used to define a maximum lens-source impact parameter magnitude that would give rise to detectable astrometric signal amplitude \citep{Dominik2000},
\begin{equation}
    u_{T} = \frac{\theta_{E}}{\delta_{T}}\,.
\end{equation}
Neglecting lens-source parallax motion, the duration of an astrometric microlensing event is the time a source spends detectable within $u_{T}$ \citep{Honma2001},
\begin{equation}
t_{\text{ast}} = 2t_{E}\sqrt{u^{2}_{T}-{u_{0}}^{2}}\,.
\label{eq:tast}
\end{equation}
If $u_{T}<u_{0}$, $t_{\rm ast}$ is unphysical. For a given observational time baseline, $T_{\text{obs}}$, and dense enough observing cadence, events with $t_{0}$ within $T_{\text{obs}}$ and with $u_{0}<u_{T}$ will be detectable so long as $t_{\text{ast}}<T_{\text{obs}}$, (i.e., the event peaks and returns to baseline within the observation time). However, $t_{\text{ast}}$ can be on the order of years \citep[e.g.,][]{Belokurov2002} and longer than $T_{\rm obs}$. For an event which has $t_{0}$ within $T_{\text{obs}}$, if $t_{\text{ast}}>T_{\text{obs}}$ the amplitude of the astrometric signal is no longer necessarily representative of the signal seen by an observer. In this regime, an observer may only see a small segment of the full event within $T_{\text{obs}}$. The threshold for the closest approach between the lens and source that would give rise to a change in astrometric signal greater than $\delta_{T}$ within $T_{\text{obs}}$ is given by \citep{Dominik2000}\footnote{Eq. (\ref{eq:u_delta}) is valid when $T_{\rm obs}|\boldsymbol{\mu}_{\rm rel}|/\theta_{E}<<4 u_{T}$, i.e. the change in lens-source separation during $T_{\rm obs}$ is much less than the astrometric detectability radius or equivalently, when $t_{\rm ast}$, averaged over $u_{0}$, is $>>T_{\rm obs}$ \citep{Dominik2000}.}, 
\begin{equation}
    u_{\Delta} \approx \sqrt{\frac{T_{\text{obs}}\theta_{E}}{\delta_{T}t_{E}}}\,.
    \label{eq:u_delta}
\end{equation}
Fig.~\ref{fig:ast_signals} shows example events and these detectability criteria. In addition to dark PBH lenses, we will also consider the astrophysical lens population. For luminous lenses (Stars and WDs), flux from the lens acts to reduce the size of the astrometric microlensing signal \citep[altering the form of Eq.~\eqref{eq:full_shift}, see;][]{Bramich2018} which reduces the detectability radius around the lens to \citep{Dominik2000},
\begin{equation}
    u^{\rm lum}_{T} = \frac{u_{T}}{1+g}, \quad u^{\rm lum}_{\Delta} = \frac{u_{\Delta}}{\sqrt{1+g}}.
\end{equation}
Here, $g$ is the lens flux divided by the source flux. The astrometric microlensing signal is further reduced due to unresolved blending from unrelated sources along the line of sight, independent of the lens's luminosity. In this case the blended sources will contribute to the source centroid position \citep[see e.g., Eq.~(12) in][]{Lam2020}. The amount of blending in an event can be quantified using the blend fraction, $f_{\rm bl}$, which is the fraction of the unlensed source flux to the total blend flux including the lens and neighbouring sources. 

\section{PBH dark matter Simulation}\label{sec:simulation}

For the microlensing simulations, we used Population Synthesis for Compact Lensing Events \citep[\texttt{PopSyCLE};][]{Lam2020} with the PBH population support of \cite[][;submitted]{Pruett2022}. \texttt{PopSyCLE} allows for the simulation of a microlensing survey given a model of the Galaxy. Next, we briefly summarize its main components.  

\subsection{Galactic model and stellar evolution}

\texttt{PopSyCLE} uses a modified \citep[see Appendix A \& B of][]{Lam2020} version of \texttt{Galaxia} \citep{Sharma2011} to create a stellar model of the Milky Way based on the the Besançon model \citep{Robin2004}. Compact objects are generated via the Stellar Population Interface for Stellar Evolution and Atmospheres code \citep[\texttt{SPISEA};][]{Hosek2020}. \texttt{SPISEA} generates SOBHs, NSs, and WDs by evolving clusters matching each subpopulation of stars generated by \texttt{Galaxia} (thin and thick disk, bulge, stellar halo), assuming they are single-age, and single-metallicity populations and then injects the resulting compact objects into the simulation. \texttt{SPISEA} uses an initial mass function, stellar multiplicity, extinction law, metallicity-dependent stellar evolution, and an initial final mass relation \citep[IFMR; see e.g.,][]{Rose2022}. Separate IMFRs are used for NSs and SOBHs \citep[see Apendix C of][]{Lam2020} and WDs \citep{Kalirai2008}. SOBHs and NSs are assigned initial kick velocities from their progenitors. All values and relationships adopted for our simulations are in Table \ref{tab:simulation}. 

\begin{table*}
\begin{center}
\begin{tabular}{lll}
    \hline\hline 
    Parameter & Description & Value \\\hline
    $\rho_{0}$ & Characteristic central density of the dark matter halo & $0.0093M_{\odot}$pc$^{-3}$ \citep{McMillan2017} \\ 
    $\gamma$ & Slope parameter of the dark matter density profile  & 1 (NFW; \citealt{Navarro1996})\\
    $r_{s}$ & Milky Way dark matter halo scale radius & $18.6$kpc \\
    $m_{\text{PBH}}$ & PBH mass & ($10^{-4}$ $10^{-3}$, $10^{-2}$, $10^{-1}$, $1$, $10^{1}$, $30$, $10^{2}$, $10^{3}$) $M_{\odot}$ \\ 
    $v_{\text{esc}}$ & Milky Way escape velocity & $550$kms$^{-1}$ \citep{Piffl2014} \\
    $r_{gc}$ & Sun-Galactic center distance & $8.3$kpc \\
    $v_{\mathrm{kick, BH}}$ & Peak of initial SOBH progenitor kick distribution & $100$kms$^{-1}$ \\
    $v_{\mathrm{kick, NS}}$ & Peak of initial NS progenitor kick distribution & $350$kms$^{-1}$ \\
    IFMR & Initial-Final Mass Relation & \citet{Raithel2018} \\
    - & Extinction Law & \citet{Damineli2016} \\
    - & Bar dimensions (radius, major axis, minor axis, height) & (2.54, 0.70, 0.424, 0.424) kpc \\
    - & Bar angle (Sun–Galactic center, 2nd, 3rd) & (62.0, 3.5, 91.3) $^{\circ}$ \\
    - & Bulge velocity dispersion (radial, azimuthal, z) & (100, 100, 100) kms$^{-1}$ \\
    - & Bar patternspeed & 40.00kms$^{-1}$ kpc$^{-1}$ \\
    \hline
\end{tabular}
\end{center}
\caption{\label{tab:simulation} Summary of \texttt{PopSyCLE} simulation and PBH dark matter parameters. The galactic parameters (starting with the IFMR and down to the bulge pattern speed) were chosen to be consistent with the ``v3'' version in \citet[App.~A]{Lam2020}, as this galactic model best matched the event rates reported by OGLE~\citep{Mroz2019}.}
\end{table*}

\subsection{PBH population}

\noindent Following \cite{Pruett2022}, we assume a PBH dark matter halo density profile \citep{McMillan2017},
\begin{equation}
    \label{eq:rho}
    \rho \left(r\right) = \frac{\rho_{0}}{\left(\frac{r}{r_s}\right)^{\gamma}\left(1 + \frac{r}{r_s}\right)^{\left(3-\gamma\right)}}\,,
\end{equation}
Here, $\rho_{0}$ is the characteristic density of Milky Way dark matter halo at the Galactic center, $r_{s}$ is the Milky Way dark matter halo scale radius, and $\gamma$ is the inner slope of the Milky Way halo. Values for parameters are in Table~\ref{tab:simulation}.
Under the monochromatic mass spectrum assumption, we calculate the number of PBHs of mass $m_{\rm PBH}$ to be injected given a particular line-of-sight dark matter mass, $M_{\rm LOS}$, to $16.6kpc$ ($\approx2$ time the distance to the Bulge),
\begin{equation}
    \label{eq:num_pbh}
    N_{\text{PBH}} = f_{\text{DM}} \left(\frac{M_{\text{LOS}}}{m_{\text{PBH}}} \right)\,.
\end{equation}
$f_{\text{DM}}$ is the fraction of dark matter comprised of PBHs. Eq.~\eqref{eq:num_pbh} shows that the number of PBHs needed to make a fixed fraction of DM increases with decreasing mass.

PBHs are assigned a mean velocity, $\bar{v}$, according to an Eddington inversion model \citep{Lacroix2018}. 
$\bar{v}$ then defines a Maxwellian distribution, which the RMS PBH velocity, $v$, is sampled with a random direction from, 
\begin{equation}
    \label{eq:f}
    f(v) = \sqrt{\frac{2}{\pi}} \frac{v^{2} e^{-\frac{v^{2}}{2a^{2}}}}{a^{3}},
\end{equation}
where, $a=\bar{v}\sqrt{\pi/8}$ and $\bar{v}$ are restricted to be less than the Milky Way escape velocity \citep[$v_{\text{esc}}<550$kms$^{-1}$;][]{Piffl2014}. 
This procedure allows for fast sampling of PBH velocities, but neglects correlations between PBH mass, location, and velocity \citep{Pruett2022}. 

We investigate PBH populations with a monochromatic mass spectrum ranging $m_{\text{PBH}}$ = [$10^{-4}$, $10^{-3}$, $10^{-2}$, $10^{-1}$, $1$, $10^{1}$, $30$, $10^{2}$, $10^{3}$]$M_{\odot}$. 
This range captures the space of PBH mass likely to produce detectable astrometric microlensing events but that has not been completely ruled out \citep[e.g.,][]{Bird2023} with specific attention given to $30M_{\odot}$, which is consistent with the population model for black holes as inferred via gravitational wave observations~\citep[e.g.,][]{Abbott2020,Abbott2021,Farah2023}. To reduce Poisson noise in all simulations, we chose $f_{\text{DM}}$ such that ${\gg} 1$ detectable PBH events are generated, but not so many that the simulation becomes computationally infeasible. 
In the case of $m_{\text{PBH}} = 10^{-3} M_{\odot}$ alone, the simulations for each field had to be run multiple times with different random seeds to generate $\mathcal{O}(10)$ detectable events in total.
The final numbers shown below were then scaled down by a factor of $5$ to compensate.
Predicted numbers of detectable events can then be re-scaled as a function of $f_{\rm DM}$. 

\subsection{Galactic Bulge Time Domain Survey}\label{sec:GBTDS}

\begin{figure}
    \centering
    \includegraphics{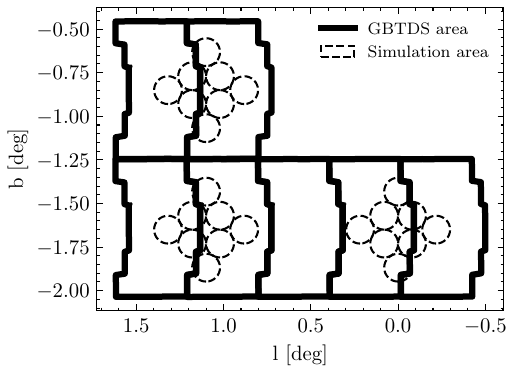}
    \caption{Simulation area and the GBTDS. Simulations are centered in three different locations - $(l,b)=(0.0^{\circ}, -1.65^{\circ}), (1.1^{\circ}, -1.65^{\circ}), (1.1^{\circ}, -0.85^{\circ})$. Each of the fields was further sub-divided into eight segments to allow efficient parallelization during the simulation runs. Individual segments required between $432$ and $720$ Central Processing Unit (CPU)-hours to compute with run-times varying with stellar densities and PBH abundance. Each group of eight segments total $0.16$ deg$^{2}$ of simulation area compared to the $1.97$ deg $^{2}$ GBTDS footprint.}
    \label{fig:sim_area}
\end{figure}

\begin{figure}
    \centering
    \includegraphics{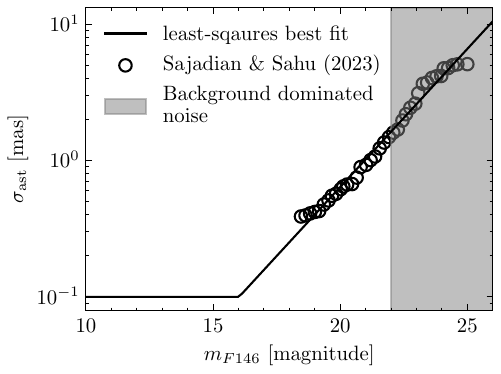}
    \caption{Estimated RST single exposure astrometric precision ($\sigma_{\text{ast}}$) as a function of F146 magnitude ($m_{\text{F146}})$. The simulated RST precision data from \cite{Sajadian2023} is fit linearly in log-space by minimizing least-squares and imposing a floor of $0.1$ mas. The shaded region indicates the magnitude range where background sources of noise become dominant \citep{Wilson2023}. The best fit model is given in Eq.~\eqref{eq:roman_ast}.}
    \label{fig:roman_ast_precision}
\end{figure}

We ran simulations, each of area $0.16$ deg$^{2}$, centered in three different places within the GBTDS area (see Fig.~\ref{fig:sim_area}). Detectable event numbers were computed by combining the three simulation centers and scaling results to the full GBTDS area of $1.97$ deg$^{2}$ \citep{Penny2019}.
To estimate the single-exposure astrometric precision of RST, we fit a linear model to the simulation data in \cite{Sajadian2023} via least squares \citep[\texttt{SciPy};][]{Virtanen2020} and imposed a floor consistent with RST's predicted absolute astrometric precision \citep[$0.1$ mas;][]{Sanderson2019},
\begin{equation}
    \sigma_{\text{ast}}(m_{\text{F146}}) = \text{max}(0.1,10^{0.2\times m_{\text{F146}}-4.23})\quad{}\text{mas}\,.
    \label{eq:roman_ast}
\end{equation}
Fig.~\ref{fig:roman_ast_precision} shows this model over the simulated astrometric precision data. This astrometric precision model is valid in the source dominated regime but neglects factors such as saturation, number of available reference stars, and source crowding issues which can all impact astrometric precision \citep[see e.g., Fig.~15 in][]{Hosek2015}. 
The transition from a source-dominated regime to a background-dominated noise regime is expected at $m_{F146}>22$ \citep[see e.g., Fig.~4 in][]{Wilson2023} for RST. Therefore we apply a conservative cut and assume that we cannot extract astrometry for sources with $m_{F146}>22$. 
For a given source of a simulated microlensing from \texttt{PopSyCLE}, the output Johnson-Cousins J and H band magnitudes is converted to an estimated $m_{\text{F146}}$-band AB magnitude using Eq. (11) in \cite{Bachelet2022}.

Finally, we adopt the suggested GBTDS survey cadence strategy in \cite{Penny2019} - a survey duration of five years with six 72-day observing seasons. Each of the seasons are spaced with gap of $\approx 111$ days, apart from season three and four, which have a larger gap of $\approx 841$ days. Within a season, we adopt a $15$ minute cadence for observations in the F146 band-pass over the full survey area.\footnote{\url{https://roman.gsfc.nasa.gov/galactic_bulge_time_domain_survey.html}} We denote this set of $41,472$ observing times over the GBTDS $\mathcal{T}_{\text{GBTDS}}$.

\subsection{Selecting events}\label{sec:cuts}

Using our simulations of the Galaxy with PBH dark matter, we select wide separation microlensing events \edits{which have} a detectable astrometric signal during the GBTDS, but no photometric signal ($u_{0}>2$) - a complementary set of events to those investigated in \cite{Pruett2022}. 

Before we consider candidate lens-source pairing, we make cuts on the PBH lens population injected into the simulation. 
Following \cite{Pruett2022}, we only consider PBH lenses within the LOS light cone of our simulated survey area. We also cut PBH lenses that are not capable of producing a detectable astrometric microlensing signal at $u_{0}=2$ - the closest lens-source separation considered in this work. 
Using Eq. \ref{eq:shift} we require an upper bound on the maximum signal amplitude, $\delta(u=2,D_{S}=\infty)>0.01$ mas - an astrometric threshold $10$ times better than the floor of Eq.~\eqref{eq:roman_ast} taking advantage of stacking $\approx 100$ observations GBTDS over 24 hours. Lenses not meeting this threshold cannot possibly cause a detectable event during the GBTDS. This effectively places a lens mass-distance cut that solely eliminates distant $m_{\rm PBH}=10^{-4}M_{\odot}$ lenses aiding computational tractability of the simulations.

When considering lens-source pairs, we require that $2<u_{0}<100$ and that the maximum lens-source impact parameter $u_{0}\theta_{E}<3000$ mas for a given pairing. These lens-source separation thresholds were chosen to capture approximately all the detectable events for our PBH mass ranges whilst keeping simulations computationally tractable. We also require that the background source magnitude $m_{\rm F146}<22$ in all source stars to eliminate events where background noise will make precision astrometry difficult (see Section~\ref{sec:GBTDS}). We also require that $\text{min}(\mathcal{T}_{\rm GBTDS})<t_{0}<\text{max}(\mathcal{T}_{\rm GBTDS})$, i.e. $t_{0}$ is within $\mathcal{T}_{\rm GBTDS}$.

Next, we make cuts on the detectability of the astrometric signals. 
Given that we require $t_{0}$ to be within $\mathcal{T}_{\rm GBTDS}$, events need to meet one of two criteria. 
\begin{enumerate}
    \item We require that $15 \text{ minutes}<t_{\text{ast}}\leq T_{\text{obs}}$. In words, the amplitude of the event is representative of the astrometric signal because the event will peak and return to baseline within $T_{\text{obs}}$. \edits{Events that are too short will be} missed by the $15$ minute cadence. Here, we use $\delta_{T}=\sigma_{\text{ast}}/\sqrt{96}$ to factor in performance gains of stacking $96$ GBTDS measurements per $24$ hours. 
    \item For events satisfying $t_{\text{ast}}> T_{\text{obs}}$, where the signal amplitude is no longer necessarily representative of the signal seen within $T_{\rm obs}$, we require that the event has a change in signal $>\delta_{T}=\sigma_{\text{ast}}/\sqrt{96}$ within $T_{\text{obs}}$, i.e, $u_{0}<u_{\Delta}$. 
\end{enumerate}
For luminous astrophysical lenses, we apply the equivalent constraints using $u^{\rm lum}_{T}$ and $u^{\rm lum}_{\Delta}$.
We then apply a cut that factors in the GBTDS cadence,
\begin{equation}
\Delta\delta_{\text{GBTDS}} = \text{max} |\delta(t_{i}) - \delta(t_{j})| : t_{i}, t_{j} \in \mathcal{T}_{\rm GBTDS}
\label{eq:delta_gbtds}
\end{equation}
requiring that $\Delta\delta_{\text{GBTDS}}> \delta_{T} = \sigma_{\text{ast}}/\sqrt{96}$. This requires that a change in signal above the detection threshold is seen in at least one pair of observations.

We also only select events with a small amount of blending $f_{bl}>0.8$, where the blend captures all light within a $90$ mas aperture consistent with RST's F146 point spread function's full width-half maximum. In addition to diluting the signal, blended lens light makes the functional form of the astrometric microlensing shift more complex, and unrelated source blending introduces many more parameters (neighbour fluxes and positions) into modeling the centroid position. These effects will act to reduce the constraints on the microlensing parameters containing lens information ($\theta_{E},t_{E},\pi_{E}$) making highly-blended events less useful for population inference. Moreover, highly blended events will likely be more difficult to detect in the GBTDS data stream. Selecting events with little to no blending also allows for better separation of the PBH population and astrophysical lens population which is systemically more blended due to lens light. 

Finally, for our detected sample of events we compute expected microlensing parameters constraints $\boldsymbol{\theta} = [t_{E}, \pi_{E},\theta_{E},t_{0}, u_{0}]$ by calculating the Cram\'er-Rao bound using the Fisher information matrix of the astrometric microlensing signal with elements,
\begin{equation}
    \mathcal{F}_{i,j} = \sum_{t\in\mathcal{T}_{\text{GBTDS}}}\frac{1}{\sigma^{2}_{\text{ast}}}\frac{\partial \delta(t)}{\partial \boldsymbol{\theta}_{i}}\frac{\partial \delta(t)}{\partial \boldsymbol{\theta}_{j}}\,.
    \label{eq:fisher}
\end{equation}
For $\delta$, we use full expression in Eq.~\eqref{eq:full_shift}, and we have assumed a white Gaussian noise model.
The diagonal elements of $\mathcal{F}^{-1}$ give a lower bound on the possible constraints of the microlensing parameters \citep[e.g., see][for example uses in microlensing]{Abrams2023}. In Eq.~\eqref{eq:fisher}, by not including the baseline source astrometry in the Fisher matrix (source reference position, proper motion and parallax), we are assuming they are measured perfectly. This is reasonable because baseline source astrometry is unlikely to dominate the error budget - end of mission relative astrometry is expected at the $3-10\mu$as precision level for $m_{\rm F146}\sim21$ sources \citep{Sanderson2019}, $10\times$ better than the floor of single exposure astrometric precision.
Moreover, source baseline astrometry can be improved by taking data after the astrometric microlensing event or by using archival baseline measurements \citep[e.g.,][]{Smith2018}.

\begingroup
\setlength{\tabcolsep}{3pt} 
\begin{table*}[t]
\begin{center}
\begin{tabular}{l|lllllllll}
    \hline\hline 
     & \multicolumn{9}{c}{\textbf{Number of Lenses} ($f_{\rm DM}$=1)} \\
 
     Event Selection Criteria & \multicolumn{9}{c}{PBH mass ($M_{\odot}$)} \\
     & $10^{-4}$ & $10^{-3}$ & $10^{-2}$& $10^{-1}$ & $1$ & $10^{1}$ & $30$  &$10^{2}$ & $10^{3}$ \\
     \hline
     \hspace{-15mm}\makecell[c]{$N_{\rm PBH}$ }& $2.1 \times 10^{12}$ & $2.1\times10^{11}$ & $2.1\times10^{10}$ & $2.1\times10^{9}$ & $2.1\times10^{8}$ & $2.1\times10^{7}$ & $6.8\times10^{6}$  & $2.1\times10^{6}$ & $2.1\times10^{5}$ \\
     \hspace{-15mm}\makecell[c]{$\delta(u=2, D_{S}=\infty)$\\$>\text{floor}(\sigma_{\text{ast}}/10)$ mas }& $4.3 \times 10^{10}$ & $2.1\times10^{11}$ & $2.1\times10^{10}$ & $2.1\times10^{9}$ & $2.5\times10^{8}$ & $2.0\times10^{7}$ & $6.8\times10^{6}$  & $2.1\times10^{6}$ & $2.1\times10^{5}$ \\
     \hspace{-15mm}\makecell[c]{Within LOS light cone} & $2.4\times10^{8}$ & $5.4\times10^{10}$ & $5.4\times10^{9}$ & $5.3\times10^{8}$ & $5.3 \times 10^{7}$ & $5.3 \times 10^{6}$ & $1.7 \times 10^{6}$  & $5.3 \times 10^{5}$ & $5.3 \times 10^{4}$ \\
     \hline
     \hspace{-15mm}\makecell[c]{($\theta_{E} u_{0}< 3000$ mas)  \\and\\ ($m_{\rm F146}<22$) \\and\\ ($t_0$ within $T_{\rm obs}$)}  &  282728 & 1898053 & 1679933 & 548976 & 163378 & 33395 & 12759 & 4032 & 407  \\
     \hline
     \hspace{-15mm}\makecell[c]{source $2 < u_{0}<100$  }& 138645 & 932655 & 823763 & 269076 & 79556 & 16195 & 6041 & 1848 & 147 \\ 
     \hline
     \hspace{-15mm}\makecell[c]{(15 minutes$<t_{\text{ast}}<$5 years) \\or\\($t_{\text{ast}}\geq$5 years and $u_{0}<u_{\Delta}$)} & 43 & 164 & 1182 & 3296 & 5583 & 4705 & 2973 & 1303 & 139 \\
     \hline
     \hspace{-15mm}\makecell[c]{$\Delta\delta_{\text{GBTDS}}>\sigma_{\text{ast}}/\sqrt{96}$ }& 11 & 32 & 344 & 1451 & 2944 & 2352 & 1437 & 702 & 101 \\
     \hspace{-15mm}\makecell[c]{$f_{bl}>0.8$}& 11 & 32 & 344 & 1410 & 2773 & 2145 & 1437 & 640 & 89 \\ 
    \hline
\end{tabular}
\end{center}
\caption{\label{tab:cuts} Astrometric microlensing event selection criteria. \edits{The} number of lenses for each PBH population remaining is shown for each cut. \edits{The first row ($N_{\text{PBH}}$) represents the total number of PBHs in the field of view before any cuts.} See Section~\ref{sec:cuts} for an explanation of each cut. All the numbers are scaled to the GBTDS area. 
}
\end{table*}


\begin{table}[t]
\begin{center}
\begin{tabular}{p{45mm}|c}
    \hline\hline 
     \hspace{-15mm}\makecell[c]{Event Selection \\Criteria} & \hspace{-10mm}\makecell[c]{Number of \\Astrophysical Lenses} \\
     \hline
     \hspace{-15mm}\makecell[c]{($\theta_{E} u_{0}< 3000$ mas) \\and\\ ($m_{\rm F146}<22$)  \\and\\ ($t_0$ and within $T_{\rm obs}$)} & 981733 \\ 
     \hline
     \hspace{-15mm}\makecell[c]{$2 < u_{0} <100$} & 480417 \\
     \hline
    \hspace{-15mm}\makecell{(15 minutes $< t_{\text{ast}} <$ 5 years) \\or\\ ($t_{\text{ast}}\geq$ 5 years and $u_{0}<u^{\rm lum}_{\Delta}$) }& 8269 \\
    \hline
    \hspace{-15mm}\makecell[c]{$\Delta\delta_{\text{GBTDS}}>\sigma_{\text{ast}}/\sqrt{96}$} & 4506 \\
    \hspace{-15mm}\makecell[c]{$f_{bl}>0.8$} & 3258 \\
    \hline
\end{tabular}
\caption{\label{tab:astro_cuts} Astrophysical microlensing event selection criteria. We do not apply the initial PBH-relevant cuts in Table \ref{tab:cuts} to the astrophysical lens population.}
\end{center}
\end{table}
\endgroup

\section{Results}\label{sec:results}

\subsection{Number of detectable events}\label{sec:event_rates}

\begin{figure}
    \centering
    \includegraphics[width=\columnwidth]{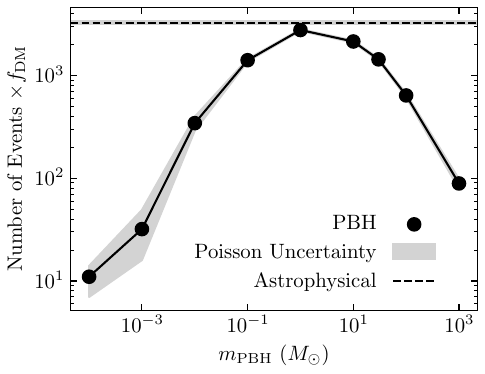}
    \caption{Predicted detectable and purely astrometric microlensing event rates ($u_{0}>2$) for the RST GBTDS scaled up to the full GBTDS area of $1.97$ deg$^{2}$ as a function of PBH mass ($m_{\text{PBH}}$). Our predicted rate of detectable and purely astrometric astrophysical (Stars, NSs, SOBHs and WDs) lensing events is plotted as a horizontal dashed line. Gray bands are derived from the Poisson uncertainties on the unscaled event numbers generated during the simulations.
    }\label{fig:event_rates}
\end{figure}

Table \ref{tab:cuts} shows number of surviving events after the cuts in Section \ref{sec:cuts} are applied successively. 
Fig.~\ref{fig:event_rates} shows the number of detectable events as a function of PBH mass. 
We find that the number of detectable events peaks for $m_{\rm PBH}=1M_{\odot}$ at $2773f_{\rm DM}$ over the GBTDS. 
The number of detectable events tapers down to $11f_{\rm DM}$ and $89f_{\rm DM}$ at $m_{\rm PBH}=10^{-4}M_{\odot}$ and $m_{\rm PBH}=10^{3}M_{\odot}$, respectively.
We find events rates of $\approx 10^{3}f_{\rm DM}$ for $m_{\rm PBH}=10^{-2}M_{\odot}-10^{3}M_{\odot}$ suggesting peak optimistic-limit GBTDS sensitivity down to $f_{\rm DM}\sim10^{-3}$ for those PBH masses. 
Fig.~\ref{fig:event_rates} also shows that the peak number of detectable events at $m_{\rm PBH}=1M_{\odot}$ is similar to the number of detectable astrophysical lens events (see Table \ref{tab:astro_cuts}).

\begin{figure*}[t!]
    \centering
    \includegraphics[width=\textwidth]{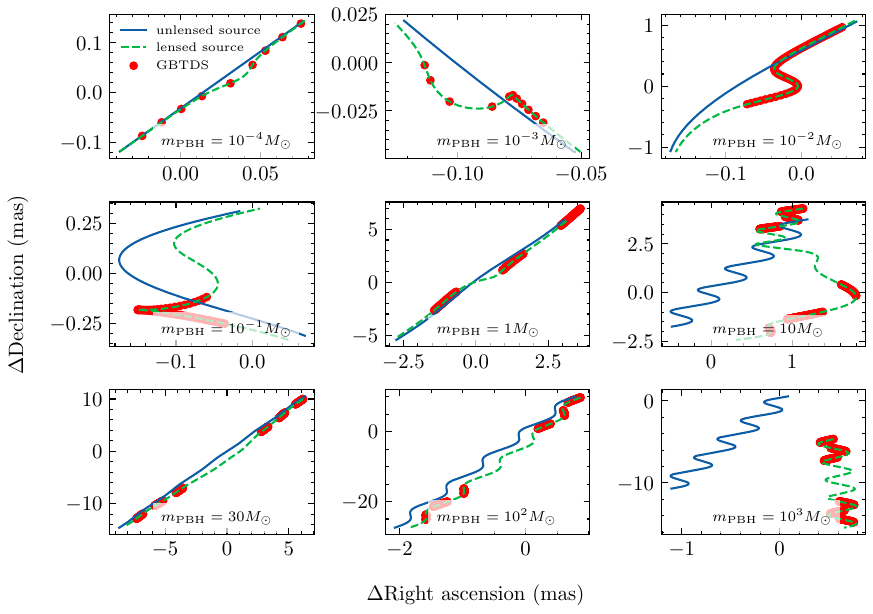}
    \caption{Examples of the astrometric microlensing deflection signal for a selection of events from the simulations passing the detectability criteria in Table \ref{tab:cuts}. \edits{Overlaid on the signals is the GBTDS stacked observing cadence of 24 hours where we have assumed that $96$ single exposures are stacked into a single measurement over each 24 hour period. The clusters of measurements seen in some panels are the GBTDS 72-day observing seasons.} 
    For the high-mass PBHs ($m_{\rm PBH}>10M_{\odot}$), events are more likely to extend beyond the five year survey and produce astrometric microlensing signals that are only partly detected. 
    Parallax motion of the lens and source was computed using Astropy \citep{astropy:2013,astropy:2018,astropy:2022} which uses values computed from NASA JPL’s Horizons Ephemeris.}
    \label{fig:example_events}
\end{figure*}

\begin{figure}
    \centering
    \includegraphics{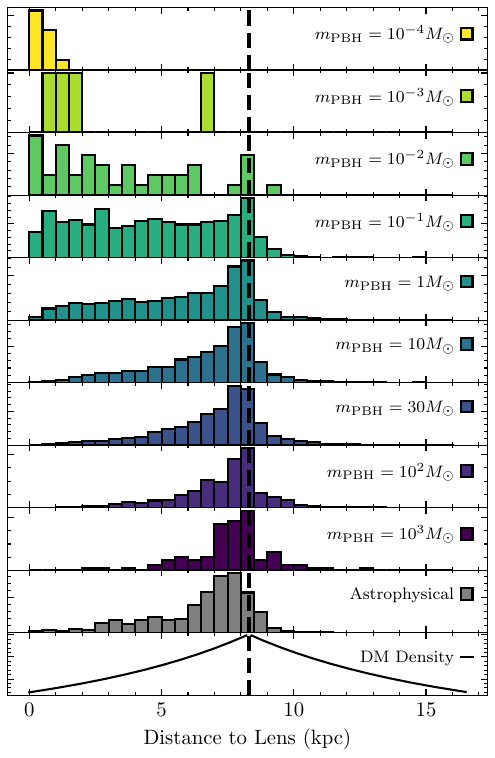}
    \caption{Distribution of distances to the PBH lenses that cause detectable astrometric microlensing events as a histogram normalised by the number of events for each $m_{\text{PBH}}$. The distance at which PBH lenses can be detected increases with mass. At $m_{\text{PBH}}\approx 1M_{\odot}$, PBHs in the Galactic Bulge are detectable which is where the bulk of the dark matter (DM) density is located. All plots are on a linear y-axis scale -  the DM density near the sun is small compared to the center of the Galaxy, but not zero.}\label{fig:lens_distances}
\end{figure}

The main reason for the decreasing event rate for $m_{\rm PBH}<M_{\odot}$ is the resulting smaller system $\theta_{E}$ ($\theta_{E}\propto\sqrt{M_{L}}$; Eq.~\ref{eq:theta_E}). 
Smaller $\theta_{E}$ values correspond to smaller $u_{T}$, decreasing the chance of a background source coming within a detectable impact parameter. 
This is further compounded by the gap in the GBTDS seasons, which are typically $>>t_{\rm ast}$ (see Fig.~\ref{fig:example_events}) for $m_{\rm PBH}<10^{-1}M_{\odot}$, meaning \edits{some} events are completely missed. 
Smaller $\theta_{E}$ also \edits{affects} the distances at which PBH lenses cause detectable events (see Fig.~\ref{fig:lens_distances}). For $m_{\text{PBH}}<M_{\odot}$, PBHs typically need to be closer than the Bulge and within $\approx7$kpc for $u_{T}$ to be sufficient to cause detectable events ($\theta_{E}\propto D^{-1/2}_{L}$; Eq. \ref{eq:theta_E}). This close lens distance bias for $m_{\rm PBH}<10^{-1}M_{\odot}$ means astrometric microlensing does not probe the bulk of the DM density near the center of the Milky Way, which is dominant over the number of PBHs increasing with decreasing $m_{\rm PBH}$ ($N_{\rm PBH}\propto m^{-1}_{\rm PBH}$; Eq. \ref{eq:num_pbh}).

For $m_{\rm PBH}>M_{\odot}$, there is a turnover in PBH event rate and it starts to decrease with increasing $m_{\rm PBH}$. 
This can be explained by larger $\theta_{E}$ causing events that are simply too slow to accumulate a detectable effect within GBTDS's $T_{\rm obs}$ of $5$ years. 
Fig.~\ref{fig:example_events} shows that for $m_{\rm PBH}>1M_{\odot}$ the astrometric events start to be detectable over the entire GBTDS survey time. The longer astrometric signals also mean that the higher $m_{\rm PBH}$ events are less \edits{affected} by the GBTDS cadence cut ($\Delta\delta_{\rm GBTDS}$; Table \ref{tab:cuts}) compared with $m_{\rm PBH}< M_{\odot}$. Fig.~\ref{fig:lens_distances} shows that for $m_{\rm PBH}>10M_{\odot}$ the lens distances are biased towards the bulk of the DM density near the Bulge. This is because at large $m_{\rm PBH}$, $D_{L}$ has to be sufficiently large to lower $\theta_{E}$ and produce a detectable change in astrometric signal within the GBTDS survey duration. The sensitivity of $m_{\rm PBH}>10M_{\odot}$ to the bulk of the DM density in the Bulge boosts \edits{event} rates somewhat offsetting the effect of $N_{\rm PBH}\propto m^{-1}_{\rm PBH}$.

\subsection{Microlensing observables}\label{sec:fisher}

\begin{figure*}
    \centering
    \includegraphics[width=\textwidth]{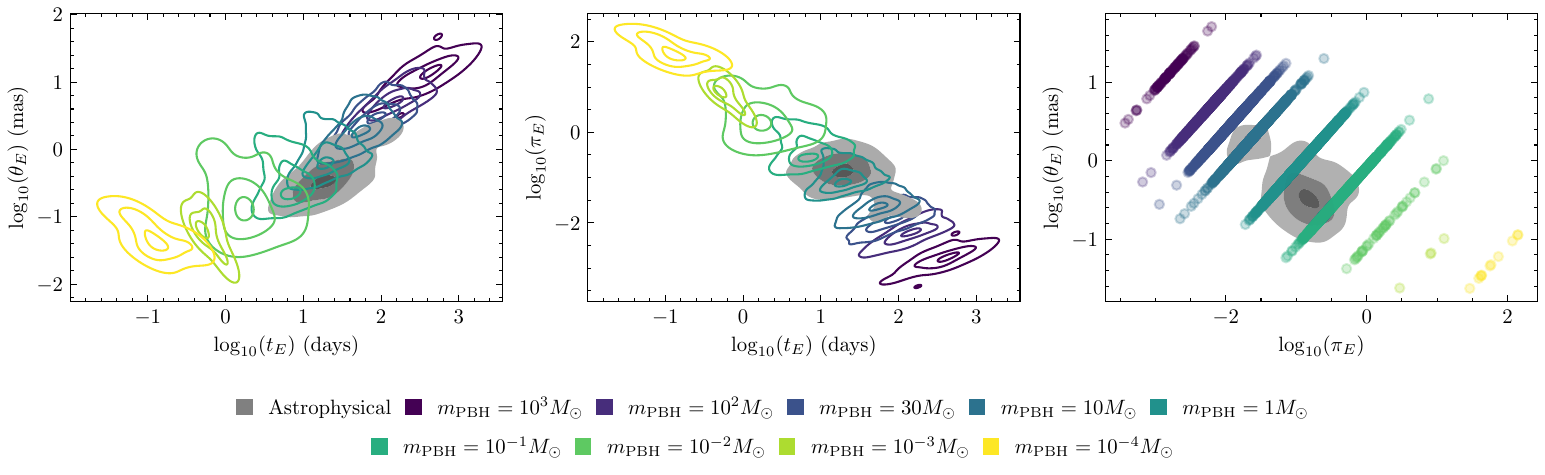}
    \caption{Positions of detectable PBH events in the intrinsic astrometric microlensing observable space ($t_{E}$, $\theta_{E}$, and $\pi_{E}$), compared with the astrophysical population of stellar, NS, WD, and SOBH lenses. In the left two panels, contours represent $0.1$, $0.5$, and $0.9$ probability and were obtained with kernel density estimation. 
    The right panel is a scatter plot for the PBHs, as the lines in $\pi_{E}-\theta_{E}$ space, which represent lenses of the same mass, are poorly represented by kernel density estimation.
    The minor mode in the $\pi_{E}-t_{E}$ and $\theta_{E}-\pi_{E}$ is due to the population of SOBHs. See Section \ref{sec:distinguishability} for discussion.}
    \label{fig:observables}
\end{figure*}

Fig.~\ref{fig:observables} shows the intrinsic distribution of three astrometric microlensing observables ($t_{E}$, $\theta_{E}$, $\pi_{E}$) for all detected events. 
The space occupied by PBHs in this space can be largely understood by $\theta_E, t_{E} \propto \sqrt{M_{L}}$, and $\pi_{E} \propto 1/\sqrt{M_{L}}$. The lines of events in the $\pi_{E}-\theta_{E}$ space are lines of constant $M_{L}$ and reflect the injected monochromatic mass PBH populations. Fig.~\ref{fig:observables} shows that both low- and high- mass PBHs lie in distinct regions of the observable space, illustrating the potential ability of these events to constrain the PBH population - a point we will investigate in Section \ref{sec:distinguishability}.

\begin{figure}
    \centering
    \includegraphics[width=\columnwidth]{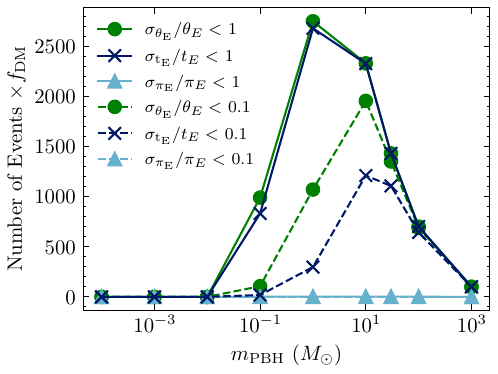}
    \caption{The number of events that satisfy the Fisher information lower-bound constraints given the characteristics of the GBTDS, i.e., the number of detectable events that can also be constrained to the respective precision. $\pi_{E}$ is not well-constrained for any $m_{\rm PBH}$. For $m_{\rm PBH}<10^{-2}M_{\odot}$, no microlensing observables are well constrained for a significant number of events. Precision constraint event rates peak for $m_{\rm PBH}=10M_{\odot}$. See Section \ref{sec:distinguishability} for discussion.}
    \label{fig:fisher}
\end{figure}

The intrinsic distribution of microlensing observables in Fig.~\ref{fig:observables} is not, however, what will be measured during the GBTDS. 
Some microlensing observables are easier to constrain than others. 
Using the Fisher information as an estimate of the lower bound on parameter constraints given GBTDS cadence and astrometric precision gives us some insight into this issue. 
Fig.~\ref{fig:fisher} shows the lower bound constraints on each microlensing parameter as a function of $m_{\rm PBH}$. 
We find that $\pi_{E}$ is not well constrained for any $m_{\rm PBH}$ and $t_{E}$ and $\theta_{E}$ are best constrained for the most events at $m_{\rm PBH}=10M_{\odot}$. 
For $m_{\rm PBH}<M_{\odot}$, we find that only a small fraction of events have well measured observables due to the short astrometric microlensing signals only being covered by a small fraction of the GBTDS observations (see Fig.~\ref{fig:example_events}). 
The overall shape of the distribution of $t_{E}$ and $\theta_{E}$ constraints in Fig.~\ref{fig:fisher} mirrors the rates in Fig.~\ref{fig:event_rates}.

The lack of constraint for $\pi_{E}$ for any $m_{\rm PBH}$ is not surprising. 
At $|\boldsymbol{u}|>2$, deviations in the relative lens-source trajectory due to parallax motion are not detectable because the $\delta\propto |\boldsymbol{u}|^{-1}$ (Eq.~\ref{eq:shift}) and is therefore not sensitive to small trajectory changes \citep[e.g.,][]{Gould2014}. 
This is in contrast to microlensing events with photometric signals at close lens-source separations ($|\boldsymbol{u}|<2$) where $\pi_{E}$ can be constrained for some appreciable fraction of events \citep[e.g.,][]{Wyrzykowski2016,Golovich2022, Kaczmarek2022}. 
Due to the fact $\pi_{E}$ is unlikely to be well measured for our events, \edits{the next Section will address} the distinguishability of PBHs lens populations from the astrophysical lens population using the $t_{E}-\theta_{E}$ space only. Difficulty in measuring $\pi_{E}$ also suggests that these events are unlikely to yield to precise \edits{lens-mass} determinations.

\subsection{Distinguishability from Stellar lenses}\label{sec:distinguishability}

In addition to predicting the number of astrometric PBH microlensing events for the GBTDS, it also important to investigate how distinguishable PBH lenses are from the astrophysical lens population (Stars, WD, NS and SOBHs) because this will \edits{affect} the quality of PBH population constraints that can be obtained \citep{Perkins2023}.
There are a variety of tools that can quantify separability; here we will use two methods that contain complementary information. 
First, we compute a distance measure of the intrinsic observable $t_{E}-\theta_{E}$ distributions for our GBTDS detectable events considering two population models: one with the astrophysical and PBH populations and one with only the astrophysical population.
Secondly, we compute \edits{the} expected rates of seeing information-rich, ``golden'' or ``unique'' events in regions of $t_{E}-\theta_{E}$ space that are not occupied by astrophysical lenses which could provide strong evidence for a PBH population.
The former analysis focuses on bulk properties of the lens distributions and if PBHs cause significant perturbations to those properties, while the second aims to quantify if a PBH population can cause unique small-scale signatures in the $t_{E}-\theta_{E}$ space unexplainable by an astrophysical population. 

Fig.~\ref{fig:observables} shows that as $m_{\rm PBH}$ diverges from the astrophysical mass ranges, the PBH population models become better separated from the astrophysical population, as expected. 
It is important to note at this point, however, that our simulations via \texttt{PopSyCLE} of the astrophysical lenses do not contain substellar objects such as brown dwarfs and free-floating planets. 
This means that the following analysis is likely to over-estimate the distinguishability of PBHs from the astrophysical population in the ${<}10^{-1}M_{\odot}$ mass ranges. 

As shown in Sec.~\ref{sec:fisher}, $\pi_{E}$ is unlikely to be well measured for our astrometric events so we focus our attention on $t_{E}-\theta_{E}$ space to investigate separability. 
Assuming that $t_{E}$ and $\theta_{E}$ are measured perfectly, we compare how similar the intrinsic probability distributions are for an event to be produced in the $t_{E}-\theta_{E}$ space for two population models: a lens population with only astrophysical lenses, $p(t_E, \theta_E | {\rm Astro})$, and the simulation including both an astrophysical and PBH lens population, $p(t_E, \theta_E ) \equiv p({\rm PBH}) p(t_E,\theta_E | {\rm PBH}) +p({\rm Astro}) p(t_E,\theta_E | {\rm Astro})$, i.e., the probability of an event with parameters $t_E$ and $\theta_E$ marginalized over both possible lens classes. 
The priors of an event belonging to one class of the other, $p(\text{PBH})$ and $p(\text{Astro})$, are simply the relative rates normalized to one marginalized over the entire parameter space.
These two population models are utilized in favor of comparing the astrophysical-only model directly with the PBH-only model as the PBH-only model is never assumed to fully describe all of the data. 
The comparison to follow is intended to reflect the loss of information incurred by implementing the wrong model, and as such, the PBH-only model would be inappropriate as it is never assumed to be a fully viable, independent population model separate of the astrophysical-only population model.
We compare the information content of these two distributions using the Hellinger distance (using Gaussian kernel density estimation via \texttt{SciPy.stats}; \citealt{Virtanen2020}),
\begin{equation}\label{eq:hellinger}
    H^{2} = 1-\int{\sqrt{p(t_E, \theta_E  )p(t_E, \theta_E | {\rm Astro} )  }}d\theta_{E}dt_{E}\,.
\end{equation}
Here, $0< H < 1$, \edits{where} a larger Hellinger distance means that the two distributions are less similar. The Hellinger distance was used over other metrics (e.g., the Kullback–Leibler (KL) divergence), due to its symmetry under transposition of the distributions and it being bounded between $0$ and $1$. 
Furthermore, we found that computation of the KL divergence was unstable due the bulk of the probability of the PBH model for small and large masses being in the tails of the astrophysical-only population model. 

\begin{figure}
    \centering
    \includegraphics[]{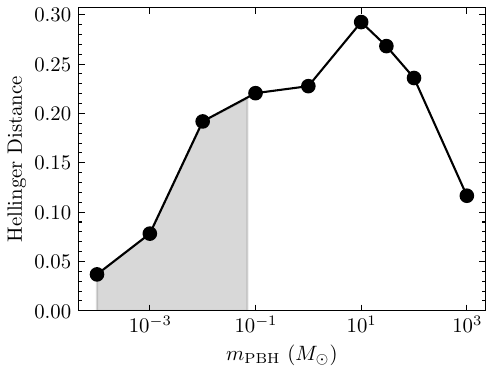}
    \caption{Hellinger distance between $t_E$-$\theta_E$ distributions of the PBH and astrophysical populations of lenses and just the astrophysical population (including stars, WDs, NSs, and SOBHs). Larger Hellinger distance means that distributions are more divergent. We see maximal divergence around $m_{\rm PBH} = 10 M_{\odot}$, achieved by balancing the relative abundance of PBHs as well as the geometric difference between the PBH and astrophysical distributions. $m_{\rm PBH}<0.07M_{\odot}$ (gray band) should be treated as an upper-bound because our simulation does not include substellar objects which could occupy the same region of $t_E$-$\theta_E$ space as PBHs.}
    \label{fig:hellinger}
\end{figure}

\begin{figure*}[t]
    \centering
    \includegraphics[width=\textwidth]{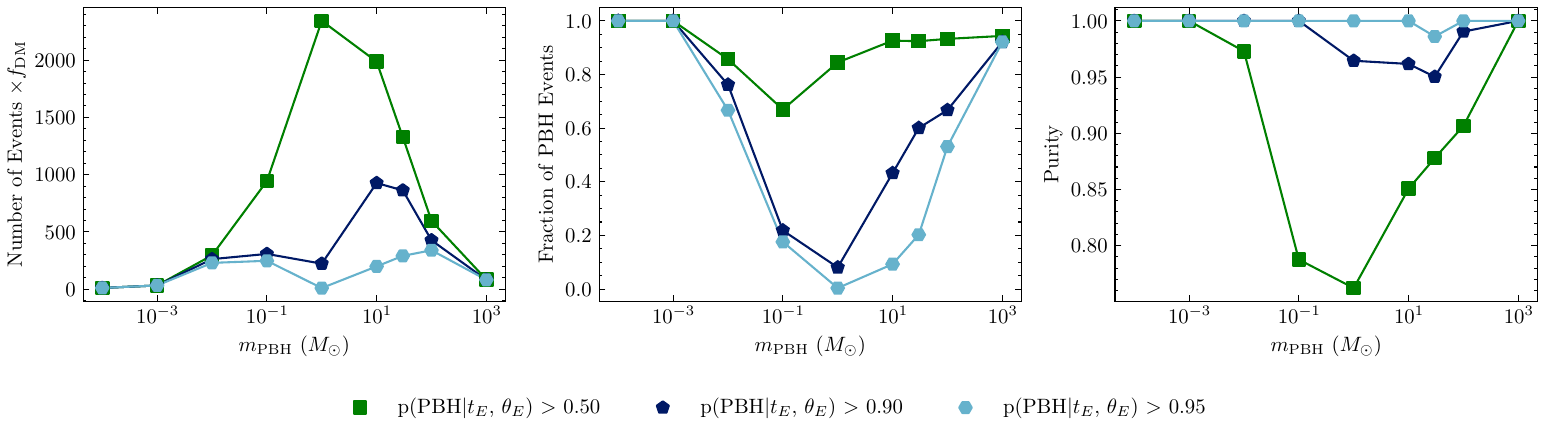}
    \caption{Statistics of PBH events that lie in distinct regions of intrinsic $t_{E}-\theta_{E}$-space from astrophysical lenses. In all plots, the probability of \edits{a} lens belonging to the PBH population given a value of $t_{E}$ and $\theta_{E}$ are shown for thresholds $0.5$, $0.9$, $0.95$. All numbers assume that $t_{E}$ and $\theta_{E}$ can be measured perfectly from astrometric observations i.e., \edits{assuming} astrometric observation noise is neglected. Left: Number of PBH events multiplied by $f_{\text{DM}}$ detectable over the GBTDS duration. 
    Middle: the fraction of PBH events that satisfy the threshold normalized by the total number of PBH events in the simulation output (independent of $f_{\text{DM}}$).
    Right: The purity of the sample, i.e., the number of PBH events that satisfy the threshold divided by the total number of events (astrophysical and PBH) that satisfy the threshold for $f_{\text{DM}}=1$.}\label{fig:mass_vs_prob}
\end{figure*}

Fig.~\ref{fig:hellinger} shows the Hellinger distance for each $m_{\rm PBH}$. 
The relationship between the Hellinger distance and $m_{\rm PBH}$ is driven by two effects: relative number of PBH to astrophysical events and the separation between the PBH and astrophysical event distributions in $t_{E}-\theta_{E}$. 
Generally, the Hellinger distance decreases with decreasing $m_{\rm PBH}$. 
Although small $m_{\rm PBH}$ PBH tend to be well-separated from the astrophysical lens population in $t_{E}-\theta_{E}$ space, the trend is dominated by the number of PBH events dropping significantly below astrophysical rates (Fig.~\ref{fig:event_rates}), meaning that the PBH perturbation to the astrophysical population becomes small. 
Fig.~\ref{fig:hellinger} shows a large decrease in the separability of PBHs from $m_{\rm PBH} = 10M_{\odot}$ to $m_{\rm PBH} = M_{\odot}$, which can be explained by the significant overlap with the bulk of the astrophysical lenses at this mass. 
Fig.~\ref{fig:hellinger} also shows a turnover in separability for $m_{\rm PBH}\approx10M_{\odot}$ which can be explained by a decreased PBH event rate (Sec.~\ref{sec:event_rates}).

To quantify how many PBHs occupy ``unique'' regions of $t_{E}-\theta_{E}$ space, we construct boundaries where the probability of an event containing a PBH lens over an astrophysical lens reaches some threshold ($0.5$, $0.9$ and $0.95$).
``Unique'' or ``golden'' events are then generally defined to be \edits{the} events which lie in those high probability contours (properly identified as events in the $0.5$, $0.9$ and $0.95$ confidence regions when relevant).
This probability is the posterior probability of a lens belonging to a class,
\begin{equation}\label{eq:PBHPosteriorProbability}
    p({\rm PBH} | t_E, \theta_E) = \frac{p({\rm PBH}) p(t_E, \theta_E | {\rm PBH } )}{p(t_E, \theta_E)}\,,
\end{equation}
where $p(t_E, \theta_E)$ is defined above.
The number of events are then simply calculated by computing the fraction of simulation samples from \texttt{PopSyCLE} which fall within the boundaries.
Even when the number of PBH events in unique regions are low, they can still provide constraining population information. 

Fig.~\ref{fig:mass_vs_prob} shows number of PBH events, fractions of PBHs events, and purity of PBH events in high-confidence regions of $t_{E}-\theta_{E}$. 
In the leftmost panel of Fig.~\ref{fig:mass_vs_prob}, the rate of events which satisfy $p({\rm PBH} | t_E, \theta_E) \geq 0.9$ generally increases with $m_{\rm PBH}$ until $m_{\text{PBH}}=10 M_{\odot}$\edits{. Although similar} to Fig.~\ref{fig:hellinger}, the dip at $1 M_{\odot}$ is due to the large overlap between PBHs and astrophysical lenses as shown in Fig.~\ref{fig:observables}.

The rise in events in ``unique'' regions can be understood as the combination of the same effects that lead to the Hellinger distance distribution (see Fig. \ref{fig:hellinger}). 
While the intrinsic number of PBHs (to explain all of DM) is highest for the lowest mass PBHs, the number of detectable events scales with the mass of the PBHs (see Fig.~\ref{fig:event_rates}). 
Secondly, the calculation depends on the confusion between the distributions.
When considering weaker uniqueness thresholds ($p({\rm PBH} | t_E, \theta_E) = 0.5$), the trend follows a similar shape as the number of events by PBH mass (Fig.~\ref{fig:event_rates}), suggesting the dominant effect is the number of detectable events rather than the overlap of the distributions.

The middle panel of Fig.~\ref{fig:mass_vs_prob} also shows the confusion of the $m_{\rm PBH}{\sim} 1 M_{\odot}$ with the astrophysical lens population.
The fraction of the PBH events that are detectable \emph{and} in the ``unique'' region sharply falls in this mass range indicating that the boundaries of the ``unique'' region do not contain the mode of the distribution $p(t_E, \theta_E | {\rm PBH})$ for this specific PBH population model.
However, the contours determined by the edge cases of the mass distributions considered here ($ M_{\rm PBH}< 0.1 $ and $M_{\rm PBH} > 50$) contain an appreciable fraction of the detectable PBH events, indicating \edits{that} the mode of the $p(t_E, \theta_E | {\rm PBH})$ distribution is reasonably separate \edits{from} the astrophysical distribution.

Finally, the third panel of Fig.~\ref{fig:mass_vs_prob} shows the purity of the region selected by our methodology which is the fraction of events in the ``unique'' region \edits{that} are truly PBH events normalized by the total number of events within these regions (all assuming $f_{\text{DM}} = 1$). 
With sufficiently strong criteria ($p({\rm PBH} | t_E , \theta_E) \geq 0.9$), the purity remains very high across the mass range. 
For weaker criteria ($p({\rm PBH} | t_E , \theta_E) \geq 0.5$), the purity drops drastically for PBHs with a mass of $1M_{\odot}$, again showing the impact of the large degree of overlap between the distributions in $t_E$-$\theta_E$ space for the PBH and astrophysical distributions.

\subsection{PBH population constraints}\label{sec:popConstraints}

\begin{figure}
    \includegraphics[width=\columnwidth]{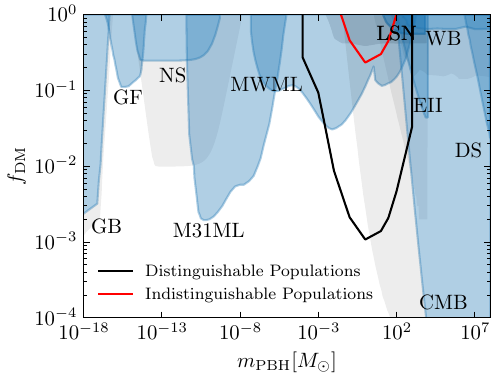}
    \caption{Projected sensitivity of the GBTDS astrometric microlensing from this work alongside current PBH constraints taken from \cite{Bird2023}. \edits{The red} line (pessimistic) is the $95\%$ confidence line on an $f_{\rm DM}$ constraint using event number information only (i.e., $t_{E}$, $\theta_{E}$, $\pi_{E}$ contain no information on the lens class) and assuming we know the number of astrophysical microlensing events to $10\%$. \edits{The black} line (optimistic) is the $95\%$ confidence line on an $f_{\rm DM}$ constraint assuming that PBH lenses can be definitively separated from the astrophysical population. Both lines are derived from the Cram\'er-Rao bound. The likely achievable constraint will be between these two lines.
    M3ML and MWML are microlensing results towards M31 \citep{Niikura2019a} and looking through the Milky Way, respectively \citep{Alcock2001, Tisserand2006, Wyrzykowski2011b, Blaineau2022}. Other constraints are supernovae lensing \citep[LSN;][]{Zumalacarregui2018},  Eridanus II dwarf galaxy \citep[EII;][]{Brandt2016,Li2017}, wide binary stars \citep[WB;][]{Quinn2009,Yoo2004}, dwarf galaxy dynamical heating \citep[DH;][]{Lu2021,Takhistov2022, Takhistov2022b}, X-ray binaries \citep[XB;][]{Inoue2017}, CMB distortions \citep[CMB;][]{Ali-Haimoud2017, Ricotti2008}, disk stability \citep[DS;][]{Xu1994}, gamma-ray background \citep[GB;][]{Carr2016}, gamma ray-femtolensing \citep[GF;][]{Carr2016dm}, and neutron star capture \citep[NS;][]{Capela2013}. Blue and grey band represent more and less conservative constraints, respectively.}
    \label{fig:population}
\end{figure}

Fig.~\ref{fig:population} shows the estimated constraints which can be derived from this work, alongside other current PBH constraints in $f_{\rm DM}-m_{\rm PBH}$.
To derive rigorous, population-level constraints would require modeling an inhomogeneous Poisson process within a hierarchical framework \citep[e.g.,][]{Perkins2023} and is beyond the scope of this work. 

To estimate the achievable constraints, we will consider two limiting cases, both of which reduce the inhomogeneous Poisson process to homogeneous Poisson statistics.
In these approximations, all information about the distributions in the event-modeling space ($t_E$, $\theta_E$, and $\pi_E$) will be neglected except in the broadest of terms as the uncertainty on the event parameters ($t_E$ and $\theta_E$) will be negligible in these limits.
Constraints using the full methods derived from inhomogeneous Poisson statistics will likely fall between these two extremes, with some $m_{\rm PBH}$ conforming more to one over the other.

The first, pessimistic approximation assumes that each of the two populations (astrophysical and PBH) are indistinguishable, in which case the inhomogeneous Poisson process reduces to two, independent Poisson processes.
This approximation would be more consistent with the PBH mass model of $m_{\text{PBH}}=1M_{\odot}$, which is maximally overlapping with the astrophysical population in $t_E$-$\theta_E$ space (see Fig.~\ref{fig:observables}).
If we assume the number of detected microlensing events is described by the sum of these two independent Poisson processes, one described by the astrophysical model with expected number of events $N^{\text{Astro}}$ and one described by the PBH model with expected number of events $f_{\text{DM}} N^{\text{PBH}}$, we can calculate the Fisher information on the parameter $f_{\rm DM}$ itself (similar to Section~\ref{sec:cuts}). 
An added complexity, however, is that the degeneracy between the two, indistinguishable Poisson processes without prior information is exact.
Of course, this degeneracy is partially broken by prior information about the astrophysical rate.
To account for this degeneracy in a realistic way, we construct a Fisher matrix in the two dimensional space spanned by $f_{\text{DM}}$ and $N^{\text{Astro}}$, but impose a prior on $N^{\text{Astro}}$.
This incorporates realistic uncertainty on \edits{the} DM fraction due to degeneracy with the astrophysical modeling and marginalizes over it.
This can then be translated to an estimate of the covariance of $f_{\text{DM}}$ through the Cram\'er-Rao bound leading to an estimated constraint on $f_{\rm DM}$ (at $95\%$ confidence) of 
\begin{equation}
f_{\text{DM}} \leqslant 1.96  \sqrt{\frac{ N^{\text{Astro}}+ \sigma_{N^{\text{Astro}}}^2}{\left(N^{\text{PBH}}\right)^2}}\,,
\end{equation}
where $\sigma_{N^{\text{Astro}}}$ is the variance on the Gaussian prior imposed on the astrophysical rate, taken to be $0.1 N^{\text{Astro}}$ as a conservative estimate (see App.~\ref{app:populationConstraints} for details on the derivation).
This prediction for the constraint includes the transformation from the $1\sigma$ estimate from the Fisher approximation to a $95\%$ confidence constraint, assuming a Gaussian posterior distribution on $f_{\text{DM}}$.
It was also derived by assuming the null hypothesis, i.e., that $f_{\rm DM}=0$.
Constraint predictions derived with this method are shown in Fig.~\ref{fig:population} as the red line. 
Our prediction on the constraints range from $f_{\rm DM} \approx 10^{-2}-1$ in the mass range we examined peaking at $f_{\rm DM} \approx 10^{-2}$ for a PBH mass of $M_{\rm PBH} = 1 M_{\odot}$.

To estimate an optimistic bound on the constraining power of the astrometric only events, we assume that the two populations (astrophysical and PBH) are completely distinguishable, i.e., that they are separable in parameter space.
The PBH mass model $m_{\text{PBH}}=10^{-4}M_{\odot}$ would be more consistent with this assumption (conditioned on the astrophysical output from \texttt{PopSyCLE}, which is lacking sub-stellar objects), as this model is maximally disparate from the astrophysical population in $t_E$-$\theta_E$ space (see Fig.~\ref{fig:observables}).
In this most optimistic scenario, where we assume that all PBH events can be exactly identified as such, we turn our number of detectable PBH events described in Section \ref{sec:event_rates} into a $95\%$ confidence bound on detecting PBHs across $m_{\rm PBH}$, with the following,
\begin{equation}
    f_{\rm DM} \leq \frac{3}{N^{\rm PBH}}\,,
    \label{eq:optimistic_constraint}
\end{equation}
where $N^{\rm PBH} $ is the expected number of PBH events from these simulations assuming $f_{\text{DM}} =1$, separated by mass. The numerator in Eq.~\eqref{eq:optimistic_constraint} originates from inverting the requirement $p(N^{\rm PBH}\geq 1 | f_{\text{DM}}) = 0.95$ for the corresponding value of $f_{\text{DM}}$, yielding a factor of $|\ln(0.05)|\approx 3$ (see Eq.~\eqref{eq:optimisticIntermediateConstraint} and Eq.~\eqref{eq:optimisticFinalConstraint} from App.~\ref{app:populationConstraints} for details).
While the Fisher information is derivable for this situation, it is ill-conditioned in the limit of $f_{\text{DM}} \rightarrow 0$.
Physically, this simply means that the uncertainty on the rate of a Poisson process with a true rate of zero is zero, i.e., that an observation of any event in the PBH region of parameter space would exclude the null hypothesis ($f_{\text{DM}} =0$).
Instead, we use this new condition, which equates to calculating the value of $f_{\text{DM}}$ at which there is a probability $p(N_{\text{PBH}} \geq 1 ) =0.95$ for seeing at least one event with RST (see App.~\ref{app:populationConstraints} for details).
In this maximally optimistic scenario, we find a peak sensitivity of the GBTDS to $f_{\rm DM}\approx3\times10^{-4}$ at $m_{\rm PBH}=1 M_{\odot}$. The sensitivity tapers to $f_{\rm DM}\approx10^{-1}$ and $f_{\rm DM}\approx10^{-2}$ at $m_{\rm PBH}=10^{-3}$ and $m_{\rm PBH}=10^{3}$, respectively. 

Between these two approximations, we predict that GBTDS will be sensitive to novel ranges of $f_{\text{DM}}$ when compared with current photometric microlensing constraints~\citep{Alcock2001, Tisserand2006,Wyrzykowski2011b,Blaineau2022} for $m_{\rm PBH}$ from $10^{-3}M_{\odot}-10^{3}M_{\odot}$ ($10^{-1}M_{\odot}-10^{3}M_{\odot}$) for optimistic (pessimistic) statistical assumptions. 
Our predicted constraint region in Fig.~\ref{fig:population} suggests that the GBTDS may be able to fill the $m_{\rm PBH}=10^{-1}M_{\odot}-10^{3}M_{\odot}$ gap between microlensing and early universe cosmic microwave background constraints \citep{Ali-Haimoud2017,Ricotti2008} down to the $f_{\rm DM}\approx10^{-2}$ level.

\section{Discussion and Conclusion}\label{sec:conclusions}

We have estimated numbers of \edits{the} wide lens-source separation, pure astrometric microlensing events caused by a PBH population and detectable during the RST's GBTDS. 
We assumed monochromatic PBH mass spectra with masses ranging from $10^{-4}M_{\odot}$ - $10^{3}M_{\odot}$. 
We find that the number of detectable PBH events peaks at $\approx10^{3}f_{\rm DM}$ for 1 $M_{\odot}$ PBHs and tapers to $\approx 10f_{\rm DM}$ and $\approx 10^{2}f_{\rm DM}$ at $10^{-4}M_{\odot}$ and $10^{3}M_{\odot}$, respectively. 
For our sample of astrometric events, we find that $t_{E}$ and $\theta_{E}$ will be the important microlensing observable space and that $10M_{\odot}$ PBHs produce the highest number of events that are distinguishable from the astrophysical lens population. 

\edits{Importantly, this will provide constraints at larger PBH masses than current microlensing surveys, especially the $1-100 M_\odot$ mass range capable of producing LIGO-detectable gravitational wave signals \cite{Bird2016}. Other constraints in this mass range are often model dependent or are relatively weak. For example, CMB constraints rely on uncertain estimates of PBH accretion rates \edits{where} the most conservative models have limited constraining power for masses $\lesssim 10^2 M_\odot$ \cite{Ali-Haimoud2017}, and gravitational wave constraints rely on currently uncertain estimates of the black hole merger rate \cite[e.g.~][]{Jedamzik2020}. Supernovae lensing constraints are much weaker than all but the most pessimistic bounds in this forecast. Strong constraints in this mass range would render it unlikely that LIGO mergers arose from relic PBHs. Conversely, detection of a large population of black holes towards the bulge would be at least suggestive. Some cosmological simulations also predict that galactic bulges contain a population of intermediate mass black holes from previous mergers \cite{DiMatteo2023} which could optimistically be detectable as a $10^3 M_\odot$ \edits{microlensing event.} }

We translated \edits{the number} of detectable events into sensitivity and constraint predictions for the GBTDS in $f_{\rm DM}-m_{\rm PBH}$ space. 
We find that the GBTDS is likely to provide competitive or novel constraints beyond current photometric microlensing surveys for PBH masses between $10^{-2}M_{\odot}-10^{3}M_{\odot}$ down to $f_{\rm DM}\approx10^{-2}-10^{-3}$ depending on the extent to which the PBH population can be disentangled from astrophysical lens population.
If realized, this predicted sensitivity of the GBTDS is likely to probe the unexplored region $f_{\rm DM}-m_{\rm PBH}$ space between current photometric microlensing surveys \citep{Alcock2001, Tisserand2006, Wyrzykowski2011b, Blaineau2022} and early universe cosmic microwave background PBH constraints \citep{Ali-Haimoud2017, Ricotti2008} down to $f_{\rm DM}\approx10^{-2}$.

At low PBH mass ($<10^{-1}M_{\odot}$), the GBTDS's sensitivity to astrometric microlensing events is limited by a combination of these short-timescale events falling in observation-season gaps and the astrometric microlensing signal only being able to probe the local ($<2$kpc) DM density. 
Moreover, we find that for PBHs $<10^{-1}M_{\odot}$, \edits{the} GBTDS cadence and precision is not sufficient to constrain any of the microlensing observables for pure astrometric events with $u_{0}>2$. 
This suggests that complimenting GBTDS astrometry with other sub-mas capable observatories (e.g., with JWST; \citealt{Gardner2006} or HST) during the survey will boost event rates and tighten the obtainable PBH population constraints. 
Large-scale strategies of filling the GBTDS season gaps or a more targeted approach of following up individual short events \citep[e.g, via the use of Target and Observation Management Platforms;][]{Coulter2023} could achieve this. 

At high PBH mass ($>10M_{\odot}$), the GBTDS's sensitivity to astrometric microlensing events is limited by the survey duration of $5$ years. Although these high PBH mass ranges probe the bulk of the DM density in the Galactic Bulge, their events tend to be too slow-varying to accumulate a detectable effect within the GBTDS. 
For these high PBH masses, only seeing a small segment of the event also means poorer constraints on microlensing observables. 
This suggests that complimenting the GBTDS with astrometric measurements before or after GBTDS, with the purpose of effectively extending the survey duration, would boost high-mass PBH event rates. 
This could be achieved by using archival astrometry (e.g., from the VVV; \citealt{Smith2018} or Gaia; \citealt{Prusti2016}) or astrometric followup after the GBTDS. \edits{For example, taking advantage of the possibility of a further 5 year extended RST mission, effectively doubling $T_{\rm obs}$ of the GBTDS, could improve constraints on $f_{\rm DM}$ for $m_{\rm PBH}\geq10M_{\odot}$ by a factor or $\sim2.8$ \citep[the astrometric opitcal depth is $\propto T^{3/2}_{\rm obs}$; see Eq. 68 in][]{Dominik2000}.}

Only a small number of astrometric observations of the GBTDS area at a sufficiently separated time baseline would be needed to increase the effective observation time and boost event rates \citep{Dominik2000}. This strategy may be particularly advantageous at providing PBH population constraints because $>10M_{\odot}$ PBHs tend be in unique areas of the microlensing observable space away from the astrophysical population. 

Across all PBH masses, we have only selected events with a background source with $m_{\rm F146}< 22$. This conservative cut was chosen to exclude the sources which are likely to be dominated by background noise effects \citep[see Fig.~4. in;][]{Wilson2023} making precision astrometry challenging. 
This cut, however, excludes the bulk background source population at $m_{\rm F146}\gtrsim24$. If astrometric processing methods could be developed to extract sub-mas astrometric precision from sources with $m_{\rm F146}> 22$, this would boost event rates for all PBH masses.

\edits{We have assumed an optimal RST astrometric precision floor for bright stars ($m_{\rm F146}\leq16$) of $0.1$ mas, however, the final astrometric precision of RST has yet to be determined. Applying our detectability cuts with a more pessimistic $0.3$ mas astrometric precision floor in the astrometric precision at $m_{F146}< 18$ \citep{Lam2023b} we find that for $m_{\rm PBH}=[10^{-4}, 10^{-3},10^{-2}, 10^{-1}, 1, 10, 30,100,1000]M_{\odot}$, $[18.2, 50.0, 57.1, 24.4, 9.1, 3.2, 1.8, 0.6, 0.0]\%$ of events that originally passed the detectability cuts now fail them. $m_{\rm PBH}\leq10^{-1}M_{\odot}$ are most effected by this less optimistic precision floor. These event attrition fractions can be explained by lens distance distributions and astrometric signal magnitudes. 

The decrease in attrition from $m_{\rm PBH}=10^{-1}M_{\odot}-10^{3}M_{\odot}$ is because $m_{\rm PBH}>1M_{\odot}$ events have typically more distant lenses (see Fig. \ref{fig:lens_distances}) and therefore \edits{have} more distant and fainter sources that are unaffected by the brighter source astrometric precision floor. Additionally, high-$m_{\rm PBH}$ events tend to have larger astrometric signals \edits{and} are unlikely to preferentially need bright sources for detectable events. 

For $m_{\rm PBH}\leq10^{-2}M_{\odot}$, Fig. \ref{fig:lens_distances} shows that this mass range is biased towards closer lenses \edits{meaning} closer, bright stars can be selected as sources. Moreover, small mass $m_{\rm PBH}$ typically have relatively small astrometric signals which \edits{are more} likely to require a bright star to be detectable. For $m_{\rm PBH}=10^{-4}M_{\odot},10^{-3}M_{\odot}$, the attrition rate calculation is likely noisy due to small number statistics ($\sim10$ events are being used for these calculations). We note that the attrition fraction for astrophysical lenses is $21.4\%$. }

The events considered in this work have only astrometric signals. 
This means that they will not be found using standard photometric microlensing event finding algorithms or triggering criteria \citep[e.g.,][]{Udalski2015, Husseiniova2021}. 
Equivalent astrometric event finding algorithms will have to be developed \citep[e.g.,][]{Chen2023} to process the GBTDS survey data to extract these events or to alert on astrometric microlensing anomalies in real-time for the purpose of triggering followup. \cite[e.g.,][]{Hodgkin2021, Hundertmark2018}.

There are several limitations and possible future avenues of research of this work. 
We derived sensitivities of the GBTDS in $f_{\rm DM}-m_{\rm PBH}$ conditioned on the assumption of a monochromatic PBH mass spectrum.
Assuming a monochromatic mass spectrum is the standard in the literature, but \edits{the} relaxation of this assumption can alter PBH constraints and change simulated sensitivities \citep[e.g.,][]{Green2016,Carr2017,Green2021}. 
Future work could explore relaxing the monochromatic mass spectrum assumptions and deriving constraints using the methods of \cite{Perkins2023}.

We have also required that an event must peak within the GBTDS survey time to be selected. 
This is likely to cut mainly long-duration, high mass PBH events which only have a detectable tail with the GBTDS, but could still be used to constrain the PBH population. 
This cut choice simplified the selection of detectable events using $t_{\rm ast}$ and allowed extraction of a well-defined sample of events which can be connected to rate and optical depth predictions \citep{Dominik2000}. 
Future work could focus on methods to select detectable events that don't peak within the GBTDS but that contain constraining PBH population information which will \edits{likely boost} the high-mass PBH event rates. 

Finally, we have only addressed PBH confusion with other astrometric microlensing events caused by the astrophysical lens population. 
We have neglected sources of confusion from other astrometric variables such as astrometric binaries \citep[e.g.,][]{Halbwachs2023}. 
This is likely to be the most problematic for long-duration high mass PBH events which have slow varying signals over the entire GBTDS. 
Future work should focus on how well astrometric microlensing can be separated from other astrometric variables and how to marginalize over that confusion to derive robust PBH population constraints. 

\section*{Acknowledgements}
\noindent \edits{We would like to thank the referee for comments that improved the clarity of the paper.} We would like to thank George Chapline, James Barbieri, and Macy Huston for useful discussions on this work. This work was performed under the auspices of the U.S. Department of Energy by Lawrence Livermore National Laboratory under Contract DE-AC52-07NA27344. The document number is \IMRELEASENO{}. This work was supported by the LLNL-LDRD Program under Project 22-ERD-037. This document was prepared as an account of work sponsored by an agency of the United States government. 
Neither the United States government nor Lawrence Livermore National Security, LLC, nor any of their employees makes any warranty, expressed or implied, or assumes any legal liability or responsibility for the accuracy, completeness, or usefulness of any information, apparatus, product, or process disclosed, or represents that its use would not infringe privately owned rights. 
Reference herein to any specific commercial product, process, or service by trade name, trademark, manufacturer, or otherwise does not necessarily constitute or imply its endorsement, recommendation, or favoring by the United States government or Lawrence Livermore National Security, LLC. 
The views and opinions of authors expressed herein do not necessarily state or reflect those of the United States government or Lawrence Livermore National Security, LLC, and shall not be used for advertising or product endorsement purposes. N.S.A. and J.R.L. acknowledge support from the National Science Foundation under grant No. 1909641 and the Heising-Simons Foundation under grant No. 2022-3542.

\software{This research has made use of NASA's Astrophysics Data System Bibliographic Services. \texttt{NumPy} \citep{Harris2020}, \texttt{SciPy} \citep{Virtanen2020}, \texttt{Matplotlib} \citep{Hunter2007}, \texttt{Singularity} \citep{Kurtzer2017, kurtzer2021}, \texttt{Docker} \citep{merkel2014docker}, \texttt{Astropy} \citep{astropy:2013,astropy:2018,astropy:2022}, \texttt{PopSyCLE} \citep{Lam2020}, \texttt{Galaxia} \citep{Sharma2011}, \texttt{SPISEA} \citep{Hosek2020}.}

\appendix

\section{Population Constraint Predictions}\label{app:populationConstraints}

In the absence of performing a computationally expensive, fully hierarchical framework to derive PBH population constraints possible with the GBTDS \citep[e.g.,][]{Perkins2023}, we instead rely on limiting cases and approximations to estimate them. 
As outlined in Section~\ref{sec:popConstraints}, we consider two limiting cases:
\begin{enumerate}
    \item The pessimistic case. The two populations are exactly degenerate, meaning that the distributions in the space of microlensing observables are identical. This equates to only utilizing the total number of events as relevant information, completely disregarding the parameters of the events.
    \item The \edits{optimistic} case. The two populations (astrophysical and PBH) are completely distinguishable in the space of observables (i.e., separate distributions in $t_E$-$\theta_E$ space). This equates to every observed event being uniquely identifiable as a PBH event or a astrophysical event. 
\end{enumerate}
In both of these cases, the full model (typically an inhomogeneous Poisson process model) can be reduced to homogeneous Poisson statistics. Even in these limiting cases, one could still use the derived astrometric model parameter uncertainties and values to improve the analysis, but this would only help to down-weight outlier events not to help disambiguate the class of different events (assuming the event posteriors are of the same or lesser order of extent as the population model distributions).

With the pessimistic case, the number of observed events can be modeled as the sum of two independent, homogeneous Poisson processes giving the probability of seeing $N$,
\begin{equation}\label{eq:poplikelihoodindistinguish}
    p(N | f_{\rm DM}, N^{\text{Astro}}, N^{\text{PBH}} ) = \frac{e^{-f_{\rm DM}N^{\rm PBH} - N^{\rm Astro} } \left( f_{\rm DM}N^{\rm PBH} + N^{\rm Astro}\right)^{N}}{N!}\,,
\end{equation}
where $N^{\text{Astro}}$ is the expected number of astrophysical events, and $N^{\text{PBH}}$ is the expected number of PBH events (at $f_{\text{DM}}=1$).
Now, the Fisher information framework can again be applied but as an approximation of the posterior on $f_{\text{DM}}$ as opposed to the astrometric modeling parameters (which is how it was first introduced in this work in Sec.~\ref{sec:cuts}).
The formal definition of the Fisher (defined here as $\Gamma$ to avoid confusion with the event parameter Fisher information, $\mathcal{F}$), 
\begin{equation}
    \Gamma_{i,j} = \sum_{N=0}^{\infty} \left[ - \partial_{i}\partial_{j} \ln p(N | f_{\rm DM},N^{\text{Astro}}, N^{\text{PBH}}) \right] p(N | f_{\rm DM},N^{\text{Astro}}, N^{\text{PBH}}) \,, 
\end{equation}
where $i$ and $j$ index the parameter vector of the model.
The sum over $N$ is the discrete form of the marginalization over the independent parameter from the definition of the Fisher information matrix.
We will utilize a two dimensional hyperparameter space spanned by the set of parameters $\{f_{\text{DM}}, N^{\text{Astro}}\}$. 
The inclusion of $N^{\text{Astro}}$ reflects the importance of the covariance between $f_{\text{DM}}$ and $N^{\text{Astro}}$.
In this limit and without prior information, the degeneracy between these two parameters will be exact. 
To address this issue, we must incorporate prior information. 
This is trivially accomplished by enforcing a normal prior on the astrophysical event rate,
\begin{equation}
\Gamma_{i,j}' =   \Gamma_{i,j} + \delta_{N^{\text{Astro}},N^{\text{Astro}}}\frac{1}{\sigma_{N^{\text{Astro}}}^2}\,,
\end{equation}
where $\Gamma_{i,j}'$ is the updated Fisher which reflects the prior information denoted by $\sigma_{N^{\text{Astro}}}$, the standard deviation on the astrophysical event rate from prior knowledge.
With this definition and our likelihood in Eq.~\eqref{eq:poplikelihoodindistinguish}, we can now calculate the Fisher. 
Inverting the full matrix and taking the square root of the diagonal element associated with $f_{\text{DM}}$ gives us an estimate of the uncertainty on this parameter (fully accounting for degeneracies with the astrophysical event rate and prior information)
\begin{equation}
    \sigma_{f_{\text{DM}}} =  \sqrt{\frac{N^{\text{Astro}}+f_{\text{DM}} N^{\text{PBH}}+\sigma_{N^{\text{Astro}}}^2}{\left(N^{\text{PBH}}\right)^2}}\,.
\end{equation}

This estimate of the $1\sigma$ constraint can be translated to a $95\%$ upper-bound on the confidence assuming a Gaussian posterior on $f_{\text{DM}}$ (consistent with the application of the Cram\'er-Rao bound)
\begin{equation}
    f_{\rm DM}^{95\%} \leqslant f_{\rm DM} + 1.96 \sigma_{f_{\rm DM}}|_{f_{\rm DM}=0} = 1.96  \sqrt{\frac{ N^{\text{Astro}}+ \sigma_{N^{\text{Astro}}}^2}{\left(N^{\text{PBH}}\right)^2}}\,,
\end{equation}
where we have assumed the null hypothesis ($f_{\text{DM}}=0$) when evaluating this uncertainty.

In the optimistic case where the two populations are perfectly distinguishable, we can again assume a homogeneous Poisson statistics to derive a likelihood for seeing $N$ events
\begin{equation}\label{eq:poplikelihooddistinguish}
    p(N | f_{\rm DM},  N^{\text{PBH}} ) = \frac{e^{-f_{\rm DM}N^{\rm PBH} } \left( f_{\rm DM}N^{\rm PBH} \right)^{N}}{N!}\,.
\end{equation}
Note that in this limit, with no confusion with the astrophysical population, the modeling can be done completely separately from the astrophysical population, neglecting correlations between the two models.

However, repeating the Fisher exercise as above, we get the following estimate for the variance on $f_{\text{DM}}$
\begin{equation}
    \sigma_{f_{\rm DM}} \geq \sqrt{\frac{ f_{\rm DM}} {N^{\rm PBH}}} \,.
\end{equation}
Unfortunately, the approximation for the uncertainty is not well behaved in the limiting case of the null hypothesis ($f_{\text{DM}}=0$).
The interpretation of this result in the current context is that the observation of a single event that is consistent with being a PBH rules out the null hypothesis. 

Therefore, we reconsider our estimate and instead seek to predict at what value of $f_{\text{DM}}$ would one expect to see at least a single event (at $95\%$ confidence).
If we start again with the Poisson distribution of Eq.~\eqref{eq:poplikelihooddistinguish} we obtain 
\begin{equation}\label{eq:optimisticIntermediateConstraint}
    p(N \geq 1 | f_{\text{DM}}, N^{\text{PBH}}) =  1 - p(N=0|f_{\text{DM}}, N^{\text{PBH}}) = 1 - e^{-f_{\text{DM}} N^{\text{PBH}}}  \,.
\end{equation}
Enforcing that this probability be equal to $0.95$ and inverting for $f_{\text{DM}}$ translates this result to a $95\%$ confident upper bound on $f_{\text{DM}}$
\begin{equation}\label{eq:optimisticFinalConstraint}
    f_{\text{DM}} \leqslant \frac{|\ln (0.05)| }{N^{\text{PBH}}} \approx \frac{3}{N^{\text{PBH}}} \,, 
\end{equation}
giving us the second prediction in Section~\ref{sec:popConstraints}.

\bibliography{refs}{}
\bibliographystyle{aasjournal}

\end{document}